\documentclass[amsmath, amssymb,reprint]{revtex4-1}

\usepackage[dvipdfmx]{graphicx}
\usepackage{color}
\usepackage{hyperref}
\usepackage{textcomp}
\usepackage{subfigure}

\usepackage[normalem]{ulem}
\usepackage{xcolor}

\begin{document}
		
\title{Synchronization of two bacterial flagella as a stochastic process}
		
\author{Jing Qin}
\affiliation{Department of Physics, Tohoku University, Sendai 980-8578, Japan}
\author{Nariya Uchida}
\email{uchida@cmpt.phys.tohoku.ac.jp}
\affiliation{Department of Physics, Tohoku University, Sendai 980-8578, Japan}

%%\author{Jing Qin, Nariya Uchida\thanks{uchida@cmpt.phys.tohoku.ac.jp}}
%%\inst{Department of Physics, Tohoku University, Sendai 980-8578, Japan}
				
\date{\today}
		
\begin{abstract}
%%\abst
{
Synchronization with noise is important for understanding biophysical processes at nano- and micro-meter scales, 
such as neuronal firing and flagellar rotations. 
To understand the energetics of these processes, stochastic thermodynamics approaches are useful.
Due to large fluctuations in a small system, ensemble averages of thermodynamic quantities are 
not sufficient to characterize the energetics of an individual sample. 
In this paper, we use a model for synchronization of bacterial flagella as an example, and develop an approximation method 
for analyzing the phase and heat dissipation in trajectories for different noise realizations. 
We describe the {temporal evolution} of the phase difference and heat dissipation as stochastic processes, 
and verify the analytical results by numerical simulations. 
}
\end{abstract}

\maketitle	
	
\section{Introduction}

Synchronization is ubiquitously seen in Nature. Since Huygens' discovery of synchronization in two pendulum clocks,
the study of synchronization has expanded across various fields~\cite{pikovsky2001universal,strogatz2004sync}. 
In biological systems, synchronization plays a critical role in maintaining vital functions, 
such as regulating circadian rhythms~\cite{reppert2002coordination}, 
coordinating brain activity~\cite{sporns2016networks}, 
and swimming using flagella and cilia~\cite{taylor1951analysis,gueron1997cilia,gueron1999energetic}.

Flagella and cilia provide a classical example of synchronization driven by hydrodynamic interactions, 
generating considerable interest in the field~\cite{golestanian2011hydrodynamic,uchida2017synchronization}. 
Powered by molecular motors, these active filaments 
create rhythmic fluid movements that propel microorganisms forward.
The power and recovery strokes of eukaryotic flagella and cilia are 
incorporated to reproduce metachronal waves through long-range hydrodynamic 
interactions~\cite{gueron1997cilia,gueron1999energetic,uchida2011generic,elgeti2013emergence,
brumley2015metachronal,meng2021conditions},
while other forms of interactions also contribute to coordinated
beating~\cite{friedrich2012flagellar,bennett2013emergent,narematsu2015ciliary,quaranta2015hydrodynamics,chakrabarti2022multiscale}.
Bacterial flagella have rotating helical bodies that are synchronized to form 
a bundle~\cite{reichert2005synchronization,reigh2012synchronization,tuatulea2022elastohydrodynamic}.
The cyclic motion of flagella and cilia are frequently represented by phase oscillators~\cite{uchida2017synchronization},
and experimentally modeled by orbiting colloidal particles~\cite{kotar2010hydrodynamic}. 
Additionally, as flagella operate at microscopic scales, thermal fluctuations and/or motor-induced active noises 
significantly influence their movement~\cite{goldstein2009noise}, sometimes causing order-disorder transitions~\cite{hsiao2016impurity}.
Therefore, it is essential to consider stochastic effects alongside deterministic dynamics.

Numerous studies have examined synchronization under the influence of 
noise~\cite{pikovsky2001universal,sakaguchi1988cooperative,teramae2004robustness,izumida2016energetics,hong2020thermodynamic}.
For instance, in Ref.~\cite{izumida2016energetics}, the authors explored a generic model of two phase oscillators, 
showing that synchronization minimizes energy dissipation associated with the odd part of the coupling,
while the even part has a contribution of either positive or negative sign
depending on the specifics of the model.
In Ref.~\cite{hong2020thermodynamic}, the authors discussed the energetics of cilia 
using the noisy Kuramoto model, and showed that synchronization reduces dissipation.
The energetics involved in synchronization are critical for understanding power efficiency in 
living systems~\cite{yang2021physical}. However, many studies focus on noise-averaged quantities, 
which may overlook significant fluctuations for each sample.
{
The rotation of individual bacterial flagella can be measured using the tethered cell 
method~\cite{silverman1974flagellar}, in which the cell body is tethered to a glass substrate by 
the flagellum and rotates, or the bead assay~\cite{ryu2000torque,sowa2005direct}, 
where a spherical bead is attached to the flagellum and rotates. The latter method has the advantage of 
imposing a smaller load on the flagellum compared to the tethered cell method, and the hydrodynamic drag force 
on the bead is easier to estimate. The load can be further reduced by modifying the hook, allowing for precise measurement 
of large speed fluctuations ~\cite{nakamura2020direct}.
}

In this paper, we examine a specific model of two bacterial flagella with cyclic trajectories influenced by white noise.
This model allows us to analytically investigate not only the statistical properties but also the single-trajectory dynamics 
without averaging. We validate our theoretical findings through numerical simulations.

\section{Model}

Following Ref.~\cite{uchida2010synchronization},
we consider flagella tethered to a substrate as active rotors that  move by pumping the fluid.
We take the $xy$-coordinates on the substrate and 
assume that a pair of rotors are fixed on it with distance $d$ along the $x$-axis,
as shown in Fig.~1.
Each rotor has a spherical bead of radius $a$, 
which is connected to the substrate by an L-shaped filament.
The filament consists of a vertical shaft of height $h$, which is the rotation axis of the rotor, 
and a horizontal arm with length $b$,
which gives the radius of the circular trajectory of the bead.
Using the polar angle $\phi_i$ of the $i$-th rotor, 
the positions of the two beads are 
represented by 
$\mathbf{r}_1=h\mathbf{e}_z+b\mathbf{n}_1$ and  $\mathbf{r}_2=d \mathbf{e}_x+h\mathbf{e}_z+b\mathbf{n}_2$,
where $\mathbf{e}_i$ ($i=x,z$) is the unit vector along the $x$- and $z$-axis, respectively,
and $\mathbf{n}_i(t)=(\mathrm{cos}\phi_i(t),\mathrm{sin}\phi_i(t),0)$ gives the direction of the rotor's arm.
The velocity of the bead reads
$\mathbf{v}_i=\frac{d}{dt}\mathbf{r}_i=b\frac{d}{dt}\mathbf{n}_i=b\frac{d\phi_i}{dt}\mathbf{t}_i$, 
where $\mathbf{t}_i=(-\mathrm{sin}\phi_i,\mathrm{cos}\phi_i.0)$ is 
the unit tangential vector of the orbit.

{
A tethered flagellum pumps the surrounding fluid by the activity of the flagellar motor,
which is modeled by a constant force exerted by the bead on the fluid.
We decompose the force $\mathbf{F}_i$ exerted by the $i$-th bead $(i=1, 2)$ as
$\mathbf{F}_i=F_r\mathbf{n}_i +  F_{\phi}\mathbf{t}_i+F_z\mathbf{e}_z$,
where $F_r$, $F_\phi$, $F_r$ are constants.
In a low Reynolds number condition, the forces generate the flow velocity field  
$\mathbf{v}(\mathbf{r})=\mathbf{G}(\mathbf{r}-\mathbf{r}_1)\cdot\mathbf{F}_1+\mathbf{G}(\mathbf{r}-\mathbf{r}_2)
\cdot\mathbf{F}_2$, where $\mathbf{G}$ is the Green function of the Stokes equation. 
The Green function under the no-slip boundary condition on the substrate 
is known as the Blake tensor~\cite{blake1971note}. 
For $\mathbf{r} = \mathbf{r}_2 - \mathbf{r}_1 \simeq d \mathbf{e}_x$ with $d \gg h, b, a$, 
we obtain $\mathbf{G(r)} \simeq 3h^2/(2\pi\eta d^3)\mathbf{e}_x \mathbf{e}_x$,
where $\eta$ is the shear viscosity. 
}

In the overdamped limit, the motion of rotors is determined 
by the balance of three forces 
the driving force,  the resistance force and the thermal fluctuation force. 
The driving force is the reaction of the force exerted by the flagellum on the fluid
and is given by $-F_{\phi} \mathbf{t}_i$ for the $i$-th rotor. 
The tangential component of the viscous resistance force is given by $F_{\rm viscous} 
=\mathbf{t}_i \cdot\zeta(\frac{d}{dt}\mathbf{r}_i-\mathbf{v}(\mathbf{r}_i))$, 
where $\zeta=6\pi\eta a$ is the resistance coefficient. 
The thermal fluctuation force is $F_{\rm thermal} = \sqrt{2k_{\rm B}T\zeta}\xi_i$, 
where $\xi_i$ is a white Gaussian noise satisfying
$\langle \xi_i(t) \xi_i(t')\rangle = \delta(t- t')$.
It is defined as $\xi_i={dW_t^{(i)}}/{dt}$ 
using the standard Wiener process $W_t^{(i)}$, and has the dimension of $t^{-\frac{1}{2}}$.
The equation of force balance $- F_{\phi} + F_{\rm viscous}^{(i)} + F_{\rm thermal}^{(i)} = 0$
gives the Langevin equation,
\begin{align}
\zeta b\frac{d\phi_1}{dt}
=& -F_{\phi}+\frac{9ah^2}{d^3}(F_{\phi} \sin \phi_2-F_r \cos \phi_2) \sin\phi_1,
\nonumber\\
&+\sqrt{2k_{\rm B}T\zeta}\xi_1     
\\
\zeta b\frac{d\phi_2}{dt}
=& -F_{\phi}+\frac{9ah^2}{d^3}(F_{\phi} \sin \phi_1-F_r \cos \phi_1)\sin \phi_2
\nonumber\\
&+\sqrt{2k_{\rm B}T\zeta}\xi_2.
\end{align}
We assume $F_\phi < 0$ so that each rotor has a positive intrinsic phase velocity.
Let $F=\sqrt{F_{\phi}^2+F_r^2}$ 
and  $\beta=\arctan \left(- F_{\phi}/F_r \right)$ $(0\leq\beta\leq {\pi}/{2})$.
As shown in Fig. 1, $\beta$ represents the tilt of the force from the radial direction,
which can be originated from the deviation of the flagellar axis from the anchoring point. 
For convenience, we introduce the two characteristic frequencies 
$\omega=F/(\zeta b)$ and $\omega_\phi= -F_\phi/(\zeta b)$ 
with the relation $\omega_{\phi}=\omega \sin \beta$.
We also define the dimensionless quantities $K={9ah^2}/{d^3}$ and $D={k_{\rm B}T}/(Fb)$ 
to express the relative strengths of hydrodynamic and thermal forces, respectively. 
Then the equations of motion become
\begin{align}
\frac{d\phi_1}{dt}
&=
\omega_{\phi}-K\omega \cos(\phi_2+\beta)\sin\phi_1+\sqrt{2D\omega}\xi_1,
\label{model1}
\\
\frac{d\phi_2}{dt}
&=\omega_{\phi}-K\omega \cos(\phi_1+\beta)\sin\phi_2+\sqrt{2D\omega}\xi_2.
\label{model2}
\end{align}
For typical flagella operating in water, we estimate the parameter values as
$a \sim b\sim h\sim1 \, \mu \mathrm{m}$, 
$d \sim 10 \, \mu\mathrm{m}$,  
$F \sim 10 \, \mathrm{pN}$,  
$k_{\rm B}T\sim 4 \times 10^{-21} \, \mathrm{J}$ 
and 
$\eta\sim10^{-3} \, \mathrm{Pa\cdot s}$. 
Thus we have $\omega\sim10$ Hz, $K\sim10^{-2}$ and $D\sim10^{-4}$. 
(Note that $\omega$ is not the flagellar rotation frequency, which is on the order of $10^2-10^3$ Hz
and is not considered in the present model.)

%%%%%%%%%%%%%%%%%%%%%%%%%%%%%%%%%%%%%%%%%
\begin{figure}
\centering
\includegraphics[width=\linewidth]{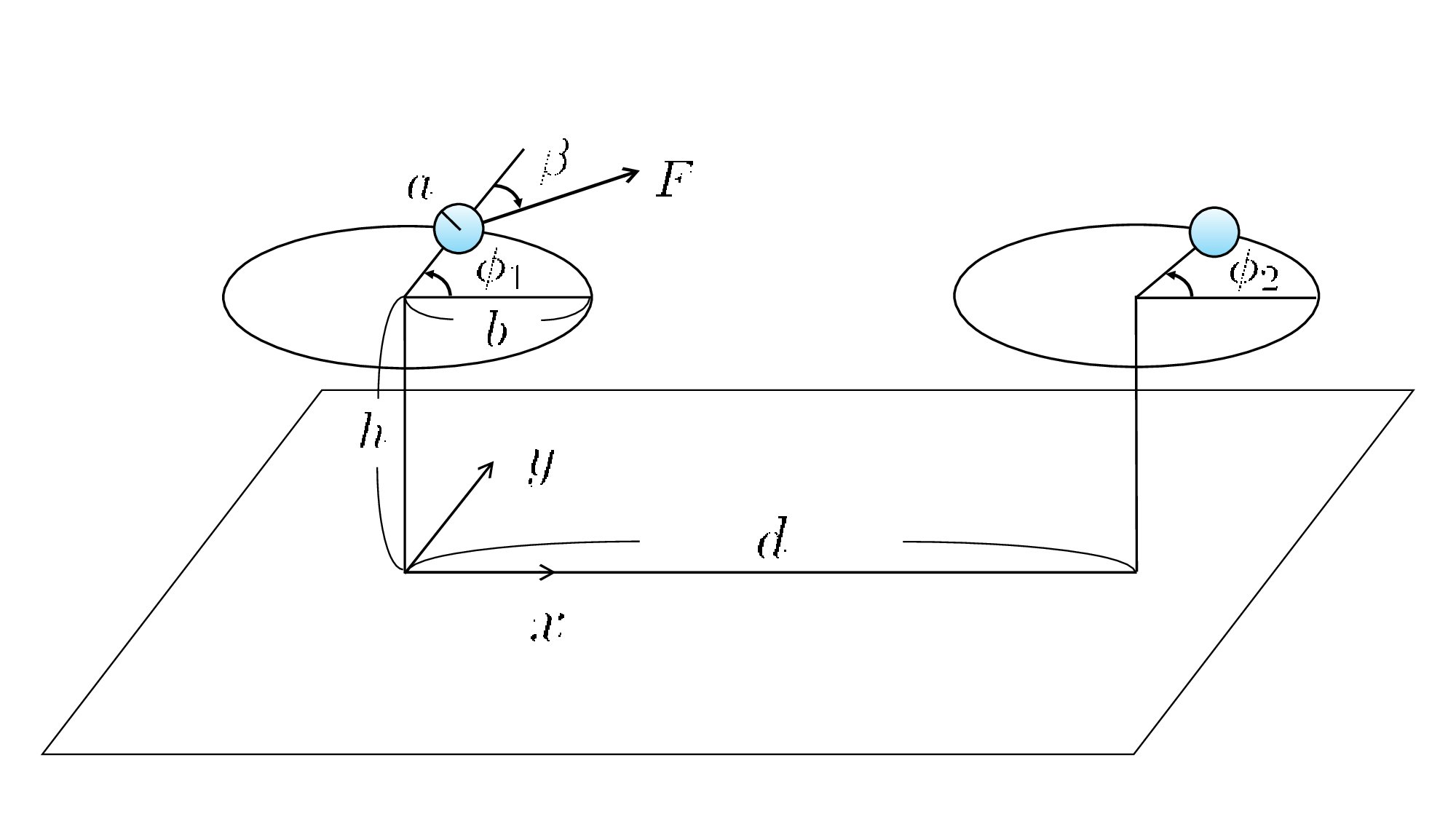}%{Fig1.png}%flagella.pdf
\caption{
(Color online) Schematic picture of the model of two flagella. Each sphere rotates on the circular trajectory of radius $b$,
exerting the force $F$, which is tilted from the radial direction by the angle $\beta$ in the clockwise direction, 
on the surrounding fluid.
}
\label{fig:1}
 \end{figure}
%%%%%%%%%%%%%%%%%%%%%%%%%%%%%%%%%%%%%%%%%%

\section{Steady state distribution}

 To analyze the system's behavior in the steady state, 
we first consider the case $\sin \beta \gg K, \sqrt{D}$, where the intrinsic phase velocity dominates 
the hydrodynamic and thermal ones (the second and third terms on the right-hand side 
of Eqs.~(\ref{model1}),(\ref{model2})), 
and introduce the slow variable $\Phi_i=\phi_i-\omega_{\phi} t$.
Using these, we rewrite Eqs.~(\ref{model1}), (\ref{model2}) as
\begin{align}
\frac{d\Phi_1}{dt} =& 
\frac{K\omega}{2} A(\Phi_2, \Phi_1)
%[\mathrm{sin}(\Phi_2-\Phi_1+\beta)-\mathrm{sin}(\Phi_2+\Phi_1+2\omega t+\beta)]
+\sqrt{2D\omega}\xi_1,
\\
\frac{d\Phi_2}{dt} =&
\frac{K\omega}{2} A(\Phi_1, \Phi_2)
%[\mathrm{sin}(\Phi_1-\Phi_2+\beta)-\mathrm{sin}(\Phi_1+\Phi_2+2\omega t+\beta)]
+\sqrt{2D\omega}\xi_2, 
\label{lan.}
\end{align}
where
\begin{align}
A(\Phi_i, \Phi_j) = &\sin(\Phi_i-\Phi_j+\beta)
\nonumber\\
& -\sin(\Phi_i+\Phi_j+2\omega t+\beta).
\end{align}
The corresponding Fokker-Planck equation for the probability distribution function $P(\Phi_1, \Phi_2; t)$ read:
\begin{align}
\frac{\partial P}{\partial t} =&
-\frac{K\omega}{2}
\left\{
\frac{\partial}{\partial\Phi_1} \left[A(\Phi_2, \Phi_1) P \right]+ 
\frac{\partial}{\partial\Phi_2} \left[A(\Phi_1, \Phi_2) P \right]
\right\}
\nonumber\\
&
+ D\omega 
\left( \frac{\partial^2}{\partial\Phi_1^2} +\frac{\partial^2}{\partial\Phi_2^2} \right) P.
\end{align}
We integrate the equation over the period ${\pi}/{\omega}$
under the slow-variable approximation that $\Phi_i$'s remain unchanged,
to get the second term of $A(\Phi_i, \Phi_j)$ vanish.
For the stationary state where ${\partial P}/{\partial t}=0$, 
we obtain
\begin{align}
0 = - \frac{K}{2} 
\biggl\{ &
\frac{\partial}{\partial\Phi_1} \left[ \sin(\Phi_2-\Phi_1+\beta) P \right]
\nonumber\\ 
&+\frac{\partial}{\partial\Phi_2} \left[ \sin(\Phi_1-\Phi_2+\beta) P \right]
\biggr\}
\nonumber\\
&+ D\left( \frac{\partial^2}{\partial\Phi_1^2} +\frac{\partial^2}{\partial\Phi_2^2} \right) P.
\end{align}
Solving the above equation, we obtain the steady-state distribution
\begin{eqnarray}
    P(\Phi_1, \Phi_2)=\frac{1}{Z}\mathrm{exp}\left[\frac{K}{2D}\mathrm{cos}\beta \, \mathrm{cos}(\Phi_2-\Phi_1)\right],
\end{eqnarray}
or equivalently 
\begin{eqnarray}
P(\phi_1, \phi_2)=\frac{1}{Z}\mathrm{exp}\left[\frac{K}{2D}\mathrm{cos}\beta \,\mathrm{cos}(\phi_2-\phi_1)\right],
\label{dist}
\end{eqnarray}
in terms of the original variables, 
where $Z$ is the normalization factor given by
\begin{eqnarray}
Z=16\pi^2I_0\left(\frac{K}{2D}\mathrm{cos}\beta\right).
\end{eqnarray}
Here, $I_0(x)$ is the 0-th order modified Bessel function.
The steady state distribution (\ref{dist}) is the von-Mises distribution for the phase difference. 
The flagella are more likely be synchronized ($\phi_1=\phi_2$)
for a stronger coupling $K$ and weaker noise $D$. 
Synchronization is also facilitated by a smaller force angle $\beta$;
for $\beta=\frac{\pi}{2}$, the distribution becomes flat and 
the system shows no tendency toward synchronization. 
For a strong coupling,
the harmonic approximation $\cos(\phi_2-\phi_1)\sim 1-\frac{1}{2}(\phi_2-\phi_1)^2$ 
gives a Gaussian distribution of the phase difference $\delta = \phi_2 - \phi_1$ 
with the standard deviation $\sqrt{2D/(K\cos\beta)}$.

The stochastic entropy for a single state is defined by $s=-k_{\rm B}\ln P$~\cite{shiraishi2023introduction}.
Using the stationary distribution (\ref{dist}),  we obtain 
\begin{eqnarray}
s(\phi_1,\phi_2)=k_{\rm B}\left[\ln Z-\frac{K}{2D}\cos\beta\cos(\phi_2-\phi_1)\right].
\end{eqnarray}
{
The stochastic entropy is minimized when the two flagella are synchronized. This implies that the synchronized state 
is the most ordered state of the system, as stochastic entropy quantifies the degree of disorder.
}
\\

\section{Heat dissipation}

In stochastic thermodynamics, for a particle satisfying the overdamped Langevin equation 
\begin{eqnarray}
\gamma \dot{x}=F(x,t)+\sqrt{2\gamma T}\xi,
\end{eqnarray}
the heat dissipation during an infinitesimal displacement $dx$ is defined as~\cite{
sekimoto2010stochastic,shiraishi2023introduction}
\begin{eqnarray}
dQ = F\circ dx =(\gamma \dot{x}-\sqrt{2\gamma T}\xi)\circ dx,
\label{dQ}
\end{eqnarray}
where $\circ$ denotes the Stratonovich product.
Applying the definition, the heat dissipation by the rotor 1 is written as
\begin{eqnarray}
dQ_1=\zeta b^2[\omega_{\phi}-K\omega\mathrm{cos}(\phi_2+\beta)\sin\phi_1]\circ d\phi_1.
\end{eqnarray}
For convenience, we normalize heat by $\zeta b^2$ and define
\begin{eqnarray}
dQ'_1=\frac{dQ_1}{\zeta b^2}=[\omega_{\phi}-K\omega\mathrm{cos}(\phi_2+\beta)\sin\phi_1]\circ d\phi_1.
\nonumber\\
\end{eqnarray}
The heat dissipation by the rotor 2 is obtained similarly.
The total heat dissipation in an arbitrary time period $0<t<T$
under the initial condition $\phi_1(0)=\phi_2(0)=0$ 
is obtained 
as, using partial integration,
\begin{align}
Q'=& \int_0^T(dQ'_1+dQ'_2)\nonumber\\
=&\, \omega_{\phi}
(\phi_1+\phi_2)
+ K \omega \left[ \cos(\phi_2+\beta)\cos\phi_1-\cos\beta \right]
\nonumber\\
&+{K\omega\int_{0}^{\phi_2}\sin\beta\cos(\phi_2-\phi_1)\circ d\phi_2}.
\end{align}

Next we compute the heat specifically by solving the stochastic 
equations (\ref{model1}),(\ref{model2}) for the phase $\phi_i(t)$. 
We change the variables to $\delta = \phi_1-\phi_2$ and $\sigma = \phi_1+\phi_2$
in Eqs.(\ref{model1}),(\ref{model2}).
For the phase difference, we obtain
\begin{eqnarray}
\frac{d\delta}{dt}=-K\omega\cos\beta\sin \delta +2\sqrt{D\omega}\xi_-,
\label{dotdelta}
\end{eqnarray}
where $\xi_- = (\xi_1 - \xi_2)/\sqrt{2}$.
Assuming that the phase difference is small, 
we approximate $\sin \delta $ by $\delta$ and solve the equation as 
\begin{align}
\delta(t) &=
2\sqrt{D\omega}e^{-K\omega\cos\beta\cdot t}\int_0^te^{K\omega\cos\beta\cdot s} dW_s
\nonumber\\
&=
2\sqrt{D\omega}\left(W_{t}-K\omega\cos\beta\int_0^t e^{K\omega\cos\beta\cdot (s-t)} 
W_s ds \right).
\nonumber\\
\label{phase difference}
\end{align}
Using this, we find $\langle\delta \rangle=\langle \phi_1-\phi_2 \rangle=0$ since the product is taken in It$\hat{\rm o}$'s sense. 
The variance reads
\begin{align}
V(t) &= \langle \delta(t)^2\rangle
\nonumber\\
&= 4D\omega e^{-2K\omega\cos\beta\cdot t}\int_0^t e^{2K\omega\cos\beta\cdot s}ds
\nonumber\\
&= 
\frac{2D}{K\cos\beta}-\frac{2D}{K\cos\beta}e^{-2K\omega\cos\beta\cdot t}.
\end{align}
In the long time limit,
it converges to the equilibrium value $2D/(K\cos\beta)$ from below.
From this, we confirm that
the small phase difference approximation is satisfied either 
when $2D/(K\cos\beta) \ll1$ or in a sufficiently short time.
{The time-evolution equation of the phase sum $\sigma$ is derived from Eqs.(\ref{model1}),(\ref{model2}) as}
\begin{align}
\frac{d \sigma}{dt}  =& 2 \omega_{\phi}-K\omega\sin(\sigma +\beta) +K\omega_{\phi}\cos{\delta}
+2\sqrt{D\omega}\xi_+,
\end{align}
{where $\xi_+ = (\xi_1 + \xi_2)/\sqrt{2}$.}
Using $\Phi_i = \phi_i - \omega_{\phi} t$
and $\Sigma = \Phi_1+\Phi_2$  
it can be rewritten as
\begin{align}
\frac{d \Sigma}{dt} =& K\omega\sin(\Sigma + \beta+ 2 \omega_{\phi} t) -K\omega\cos {\delta} \sin\beta
%\nonumber\\
+2\sqrt{D\omega}\xi_+.
\end{align}
From this and Eq.(\ref{dotdelta}),
the change of $\Phi_i$ is proportional to the dimensionless parameters $K$ and $\sqrt{D}$. 
that are in the slow variable approximation where $\sin\beta \gg K, \sqrt{D}$.
We obtain $\Sigma$ to the first order in $K$ and $\sqrt{D}$ as 
\begin{eqnarray}
\Sigma(t)&=&K\omega_{\phi}t+\frac{K\omega}{2\omega_{\phi}}\left[\cos(2\omega_{\phi}t+\beta)-\cos\beta\right]\nonumber\\
&\quad&+2\sqrt{D\omega}W_t
\label{Sigma_t}
\end{eqnarray}
Retaining only the terms in the first order of $K$ and $\sqrt{D}$ in Eq.(\ref{phase difference}),
we obtain $\delta(t) = 2 \sqrt{D\omega} W_t$.
Combining this and Eq.(\ref{Sigma_t}), and using the definition $\Phi_i = \phi_t - \omega_{\phi} t$,
we obtain
\begin{eqnarray}
\phi_i(t) &=&\omega_{\phi}\left( 1+\frac{K}{2} \right)t+\frac{K\omega}{4\omega_{\phi}}[\cos(2\omega_{\phi}t+\beta)-\cos\beta]\nonumber\\
&\quad&+\sqrt{2D\omega}W_t^{(i)}
\label{phase}
\end{eqnarray}
where $W_i$ is Wiener process and the coefficient before that is also from the addition of two noise, and
\begin{eqnarray}
Q'(t)&=&2\omega_{\phi}^2(1+K)t-2K\omega\sin(\omega_{\phi}t+\beta)\sin\omega_{\phi}t
\nonumber\\
&\quad&+2\omega_{\phi}\sqrt{D\omega}W_t
 \label{heat}
\end{eqnarray}
The first term shows the linear time dependence, the second term is periodic with the period ${\pi}/{\omega_{\phi}}$, 
and the last term is a Brownian term with increment of $\sqrt{t}$.

%%%%%%%%%%%%%%%%%%%%%%%%%%%%%%%%%%%%%%%%%%%%%
\begin{figure}[h]
\centering
\includegraphics[width=0.8\linewidth]{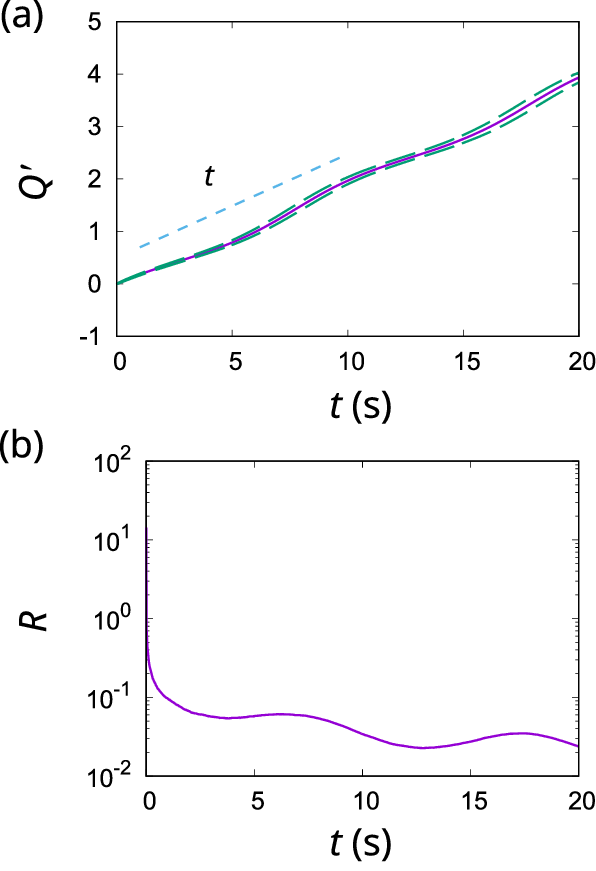}
\caption{(Color online) The case $\beta = 0.01\pi$. 
(a) Dimensionless heat versus time. The solid line shows the ensemble average $E_{Q'} = \langle Q'\rangle $ 
and the dashed lines show $E_{Q'}\pm\sigma_{Q'}$,
where $\sigma_{Q'} =\sqrt{\langle Q'^2 \rangle -\langle Q' \rangle^2}$ is the standard deviation.
The dotted line shows the slope $2\omega_\phi^2$.
(b) The relative fluctuation $R=\sigma_{Q'}/E_{Q'}$ versus time (semi-log plot).
}
\label{fig:001}
\end{figure}
%%%%%%%%%%%%%%%%%%%%%%%%%%%%%%%%%%%%%%%%%%%%%

\section{Numerical analysis}

In the analytical treatment, we used two approximations:  (i) slow variable, and (ii) small phase difference. 
The approximation (i) holds for $\sin \beta \gg K, \sqrt{D}$, while (ii) is valid if $\cos \beta \gg D/K$.
We used (i) to obtain the distribution (\ref{dist}), and (ii) for time-evolution of the phase difference 
(\ref{phase difference}), and both of them for in time-evolution of phase(\ref{phase}) and heat dissipation (\ref{heat}).
In order to verify these results and study the cases beyond the approximations, 
we performed numerical analysis by varying $\beta$ as the control parameter.
We integrated Eqs.(3),(4) for the phases $\phi_i(t)$ using the Euler-Maruyama method, 
and Eq.(15) for $Q'_1$ and a similar equation for $Q'_2$
with $\phi(t+\Delta t/2) \simeq (\phi_i(t+\Delta t)+\phi_i(t))/2$
for the Stratonovich product using the semi-implicit scheme.
The time increment $\Delta t = 10^{-4}$ is used for both sets of equations.
We took $10^3$ independent samples for the noise for statistical analysis. 
Below we focus on three representative cases: $\beta = 0.01\pi$, $0.25\pi$, and $0.49\pi$ .

For $\beta=0.01\pi$, the approximation (i) is not valid, while (ii) holds.
Comparing the first and third terms on the right hand side of Eq.(\ref{heat}),
we expect a relatively large fluctuation effect for $t < \sqrt{D\omega}/\omega_\phi \sim 1 $s. 
We plot 
the mean $E_Q(t)=\langle Q'(t)\rangle$ and
the standard deviation $\sigma_Q'(t) =\sqrt{\langle Q'(t)^2\rangle-\langle Q'(t)\rangle^2}$ 
of the heat dissipation in Fig.~\ref{fig:001}(a)
and the ratio $R={\sigma_Q}/{E_Q}$ in Fig.~\ref{fig:001}(b).
The ratio is large for short time, meeting our expectation.
In Fig.\ref{fig:001}(a), $E_Q(t)$ shows a feature of the superposition of linear increase and periodic motion. 
The slope of that is almost the same as $2\omega_{\phi}^2\approx0.2$ in (\ref{heat}). 
The curves also show a periodic part with period $T=10$ s, which
comes from the term $\sin(\omega_{\phi}t+\beta)\sin\omega_{\phi}t$.

%%%%%%%%%%%%%%%%%%%%%%%%%%%%%%%%%%%%%%%%%%%%%
\begin{figure}[h]
    \centering
    \includegraphics[width=0.8\linewidth]{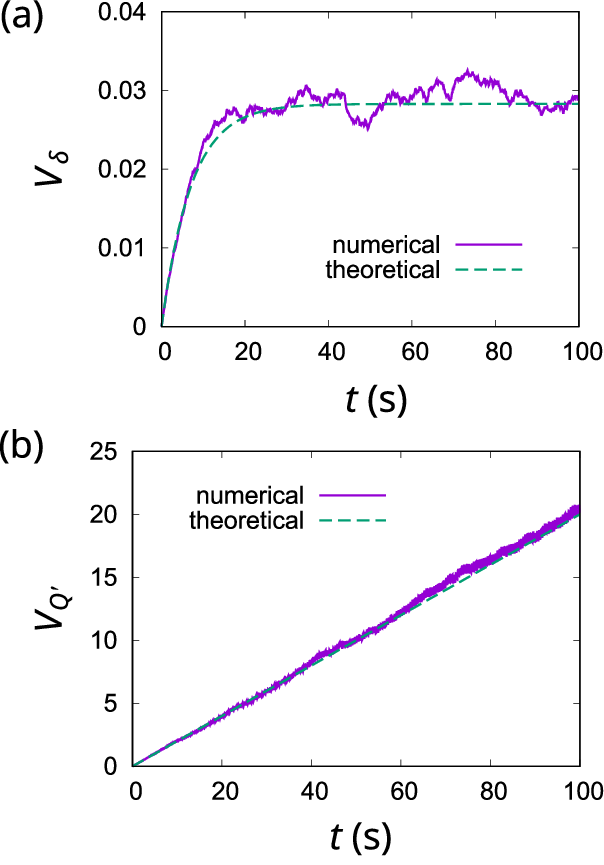}%025Q.png
    \caption{(Color online) Variation of (a) the phase difference $\delta$ and 
    (b) the dimensionless heat $Q'$ versus time for $\beta=0.25\pi$. 
    The solid lines show the numerical results, and the dashed lines show the theoretical results
    [Eqs.(\ref{phase difference}),(\ref{heat})]}
    \label{fig:025}
\end{figure}
%%%%%%%%%%%%%%%%%%%%%%%%%%%%%%%%%%%%%%%%%%%%%

%%%%%%%%%%%%%%%%%%%%%%%%%%%%%%%%%%%%%%%%%%%%%
\begin{figure}
    \centering
    \includegraphics[width=0.8\linewidth]{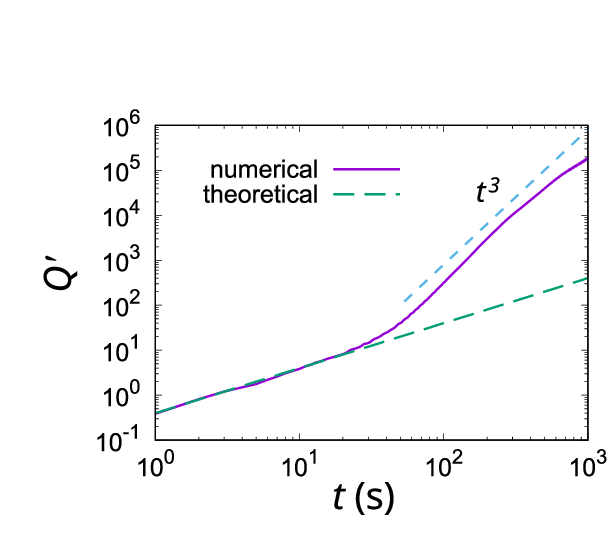}%{049Q.png}
    \caption{(Color online) Variation of the dimensionless heat for $\beta=0.49\pi$. 
    The solid line shows the numerical result, the dashed line shows the theoretical result, 
    and the dotted line shows the slope of $t^3$.}
    \label{fig:049}
\end{figure}
%%%%%%%%%%%%%%%%%%%%%%%%%%%%%%%%%%%%%%%%%%%%%

For $\beta=0.25\pi$, where both the approximations (i),(ii) are valid, 
we consider time-evolution of the variance of the phase difference
 $V_\delta(t)=\langle \delta^2\rangle=\langle(\phi_2-\phi_1)^2\rangle$ 
and the variance of the heat dissipation $V_{Q'}(t)=\langle Q'(t)^2\rangle-\langle Q'(t)\rangle^2$
in Fig.\ref{fig:025}. 
We choose the ending the time as $100\mathrm{s}$ to see both initial and steady behavior. 
The results  are in good agreement with our theoretical results; 
$V_\delta$ increases exponentially with time as in Eq.(\ref{phase difference}),
while $V_{Q'}$ linearly increases with the slope $4\omega_{\phi}^2D\omega=0.2$ 
as in Eq.(\ref{heat}).

For $\beta=0.49\pi$, the slow variable approximation holds but the small phase difference breaks down 
for a long time scale. 
We draw the {temporal evolution} of the variance of heat dissipation in Fig.~\ref{fig:049}. 
From the figure, we find that in a small time, the variance increases linearly as shown by Eq.(\ref{heat}).  
However, in the long term, the variance approximately grows with a power law with the exponent $3$,
which is not predicted by the theory.
{The crossover behavior indicates that the fluctuation effects become more significant in the weakly synchronized state at the later stage.}

{
Comparing the three cases, we find that the 
heat dissipation fluctuations at a given same time
are larger for larger $\beta$. 
This is because the tendency toward synchronization
(proportional to $\cos\beta$) decreases as $\beta$ approaches $\pi/2$, 
making the effects of noise relatively large. 
}

{\section{Discussion}}
In this paper, we proposed an analytical method to compute the motion and heat dissipation 
of interacting flagella for individual noise realizations.
We solved the phase difference and heat dissipation explicitly under the 
the slow variable and strong coupling approximations.
We compared the results with numerical simulations for three different parameter ranges,
The results show a good correspondence where the two approximations are both valid.
{
Our findings indicate that heat dissipation exhibits considerable dependence on noise realization 
and influences the energetics of bacterial flagellar synchronization. 
Experimentally, the bead trajectory obtained via a bead assay allows us to estimate the work done by 
the bead on the surrounding fluid as a function of time. Under the assumption of white noise, 
a short-time average of this work corresponds to the deterministic component, 
which is the heat dissipation (\ref{dQ}). Thus, we anticipate that our results can be compared with 
experimental data to validate the features discussed above, which we leave for future work.
}

{
This study can be extended to analyze multiple flagella, enabling us to characterize heat dissipation 
in collective synchronization. For multi-flagellated bacteria such as {\it E. coli}, flagellar bundling 
is crucial for unidirectional propulsion. In such cases, active noise originating from flagellar motors 
such as fluctuations in the number of stators could play a significant role. Although our current model 
considers only thermal noise, it can be readily extended to include active noise, 
assuming that it is additive and temporally uncorrelated.
Furthermore,} 
because the model is of a generic form, it can {be applied also to} other mesoscale systems where 
thermodynamic noise affects but does not destroy synchronization. 
\\

{\it Acknowledgment.--}
This work was supported by JSPS KAKENHI Grant Number JP24K06895 to N. U.

\bibliographystyle{jpsj}
\bibliography{qinjing2025synchronization.bib}
	
\end{document}